\documentstyle[graphics,psfig]{aa}

\begin{document}
\title{An intense and broad Fe\,K$_{\alpha}$ line observed
 in the X-ray luminous quasar Q\,0056-363 with XMM-Newton}
\author{D. Porquet\inst{1}
\and J.N. Reeves\inst{2,3}
}

\offprints{D. Porquet}
\mail{dporquet@mpe.mpg.de}

\institute{
Max-Plank-Institut f\"{u}r extraterrestrische Physik, Postfach 1312,
D-85741, Garching, Germany
\and Laboratory for High Energy Astrophysics, Code 662.0, 
NASA Goddard Space Flight Center, Greenbelt, MD 20771, USA
\and Universities Space Research Association
}

\date{Received January 20, 2003 / Accepted June 16, 2003}

\abstract{
We present an {\it XMM-Newton} observation of the radio-quiet quasar
\object{Q0056-363} (z=0.162). This is the first time that this quasar 
 is observed in the hard X-ray range (above 2\,keV). 
We find that Q0056-363 is a powerful X-ray quasar,  
with a 0.3--12\,keV unabsorbed luminosity
of about $1.2 \times 10^{45}$ erg s$^{-1}$ 
with the largest part ($\sim$ 67$\%$) 
emitted below 2\,keV.  The spectrum reveals a large featureless 
soft X-ray excess below 2\,keV and a strong broad Fe\,K$_{\alpha}$ line at 
6.4\,keV (in the quasar frame). 
The Fe\,K$_{\alpha}$ line is due to low to moderate ionization states of iron 
 (i.e., $<$ \ion{Fe}{xvii}), 
with an equivalent width of about 250\,eV and a velocity 
 width of about 25,000~km~s$^{-1}$. 
 Q0056-363 is presently the most  luminous AGN known 
to exhibit such a broad and intense Fe\,K$_{\alpha}$ line profile 
 from near neutral iron.  
The line can be fitted with a relativistic profile from an accretion disc 
around either a Schwarzschild (non-rotating)
or a Kerr (rotating) black hole. 
A combination of two thermal Comptonization components  and 
 a disc reflection model is favored to explain
 both the continuum over the energy range  0.3--12\,keV
 and the Fe\,K$_{\alpha}$ line. 
 A patchy corona covering a large part of the inner disc surface 
  is needed in order to be compatible with the accretion rate 
 inferred from the spectral energy distribution of  Q0056-363, 
 unless the mass of the black hole is much higher than 
about 5$\times$10$^{8}$\,M$_{\odot}$.    
 
}
\titlerunning{An intense and broad Fe\,K$_{\alpha}$ line 
 in Q\,0056-363}
\authorrunning{Porquet \& Reeves}
\maketitle

\section{Introduction}

In Active Galactic Nuclei (AGN), from Seyfert galaxies to quasars, 
several X-ray features can help us to understand the central region
of these powerful objects. The first one is the so-called soft excess
seen below 2--3\,keV 
(Arnaud et al. \cite{Arnaud85}; Turner \& Pounds \cite{Turner89}).
This spectral characteristic is thought to be the high energy part of the 
``big blue bump'' (BBB) extending down to 1$\mu$m, 
which contains a large fraction 
of the bolometric luminosity. Recently Pounds \& Reeves (\cite{Pounds2002}),
using {\sl XMM-Newton} observations, 
showed that a soft X-ray excess is seen in all 
their sample (6 Seyfert galaxies) 
with an amplitude and a breadth increasing with luminosity. 
Current interpretations of the soft excess range from intrinsic thermal 
emission from the accretion disc, to reprocessing of harder radiation absorbed 
in the thin disk (Pounds \& Reeves \cite{Pounds2002}).
Another important component is emission and/or absorption,
mainly in the soft X-ray range, due to the  Warm Absorber 
medium supposed to be located between the  Broad Line Region 
and the Narrow Line Region 
(e.g., Reynolds \& Fabian \cite{Reynolds95}; Porquet et al. \cite{Porquet99}). 
However these absorption and/or emission features seem to be only 
seen in low-luminosity AGN, such as Seyfert galaxies. 
The Fe\,K$_{\alpha}$ line complex observed in the 6--7\,keV range is
also an important spectral tool.
Indeed the study of this fluorescent line complex 
allows us to probe dense matter from the inner disc to the molecular torus 
as found recently in several AGN 
(e.g. \object{Mrk 205}, Reeves et al. \cite{R2001};
\object{Mrk 509}, Pounds et al. \cite{Pounds2001}). 

In this paper we present an {\it XMM-Newton} observation of 
Q0056-363, a radio-quiet quasar
  ({\it z} $= 0.162$, Hewitt \& Burbidge \cite{Hewitt89}; 
 $\alpha_{J2000}$=00$^{h}$58$^{m}$37.38$^{s}$, 
 $\delta_{J2000}$=$-$36$^{\rm o}$06$^{\prime}$04.8$^{\prime\prime}$, 
 2MASS, Cutri et al. \cite{Cutri2003}) 
  detected for the first time
in the ESO B Southern Sky Survey (Monk et al. \cite{Monk86}). 
It has strong, broad permitted optical emission lines 
 (e.g.,  FWHM(H$_{\beta}$)= 4\,700$\pm$160\, km\,s$^{-1}$, Grupe et al. \cite{Grupe99}), 
which led to it being described as a Broad-Line Seyfert 1 AGN. 
Until now, this quasar was only observed in the X-ray domain by {\sl ROSAT},
i.e. below 2.4\,keV (RASS: Thomas et al. \cite{Thomas98}, 
 HRI: Grupe et al. \cite{Grupe2001}). 
We present here the first observation of this 
object in the hard X-ray range up  to 12\,keV.

\begin{table*}[!t]
\caption{Summary of the data used to build the spectral energy distribution  of Q0056-363. 
C03: Cutri et al. (\cite{Cutri2003}).  
M03: Monet et al. (\cite{Monet2003}).  
UVW2 is the OM filter used during the present {\sl XMM-Newton} observation. 
G99: Grupe et al. (\cite{Grupe99}).   
INES: Merged Log of {\sl IUE} Observations (NASA-ESA, 1999).
}
\begin{tabular}{cccc}
\hline
\hline
\noalign {\smallskip}    
 Filter            &     magnitude or flux                    &  Observation date   &     reference\\
\noalign {\smallskip}        
wavelength or energy   band  &                            &                                &                    \\                 
\noalign {\smallskip}                       
\hline
\noalign {\smallskip}                       
K$_{\rm s}$               & 12.301$\pm$0.032 mag  &  07/1999  & C03\\
\noalign {\smallskip}                       
H                        &  13.390$\pm$0.044 mag      & 07/1999 & C03\\
\noalign {\smallskip}                       
J                         &  14.228$\pm$0.037 mag    &  07/1999 & C03 \\
\noalign {\smallskip}                       
I                         &  14.69  mag                              &  02/1987           &  M03 \\
\noalign {\smallskip}                       
[6600-7820\AA]  (Red) &     (spectrum)                                 &  10/1993          &   G99      \\
\noalign {\smallskip}                       
[4640-5950\AA]  (Blue) &      (spectrum)                                  &    10/1993       &   G99      \\
\noalign {\smallskip}                       
OM    (UWV2)       &  2.11$\pm$0.04     erg\,cm$^{-2}$\,s$^{-1}$\,A$^{-1}$     &  07/2000               & this work\\
\noalign {\smallskip}                       
 IUE     (UV: [1151--1979\AA])                    &    (spectrum)               &   10/1995                                 &  INES          \\
\noalign {\smallskip}                       
 XMM (X-rays: 0.3-10\,keV)         &    (spectrum)                             &   07/2000              &this work\\
\noalign {\smallskip}                       
\hline
\hline
\end{tabular}
\label{table:sed}
\end{table*}

Note that all fit parameters are given in the quasar's rest frame, 
with values of H$_{\rm 0}$=50\,km\,s$^{-1}$\,Mpc$^{-1}$, 
and q$_{\rm 0}$=0 assumed throughout. 
  The errors quoted correspond to 90$\%$ 
confidence ranges for one interesting parameter 
($\Delta \chi^{2}$=2.71).
 Abundances are those of Anders \& Grevesse (\cite{Anders89}). 
In the following, we use the updated cross-sections for X-ray absorption by 
the interstellar medium  from Wilms et al. (\cite{Wilms2000}).

\section[]{XMM-Newton observation}

We present an archived {\sl XMM-Newton} observation of Q0056-363 
 on July 5, 2000 (ID\,0102040701; orbit 145; exposure time $\sim$\,20\,ks). 
The EPIC MOS cameras (Turner et al. \cite{Turner2001})  
 operated in the Small Window mode, 
while the EPIC PN camera (Strueder et al. \cite{Strueder2001}) 
was operated in the standard Full Frame mode. 
For operational reasons, the MOS observations are 
divided into two separate exposures of $\sim$5\,ks, 
one with the thin filter and the other one with the thick filter.
Since the  MOS data give considerably lower net counts compared 
to the PN data (thin filter), we proceed to use only the PN for the present 
 analysis. 
The data are re-processed and cleaned (net time exposure $\sim$ 14.8\,ks) using the 
{\it XMM-Newton} {\sc SAS version 5.3.3} (Science Analysis Software) package. 
Since pile-up effect is negligible according to the {\sc epatplot} SAS task, 
X-ray events corresponding to patterns 0--4 events (single and double pixels) 
 are selected for the PN.  
A low-energy cutoff is set to 300 eV. 
The source spectrum and the light curve are extracted 
using a circular region of diameter 40$^{\prime\prime}$ (to avoid the edge
of the chip) centered on the source position. 
Q0056-363 is by far the brightest X-ray source in this
30$^{\prime}$ EPIC field-of-view. 
Background spectra are taken from an annular radius center on Q0056-363, 
between 4\arcmin~ and 8\arcmin~  (excluding X-ray point sources, 
and the columns passing through the source to avoid out-of-time events). 
The {\sc xspec v11.2} software package is used for spectral 
analysis of the background-subtracted spectrum 
using the response matrices and ancillary files derived from the SAS tasks 
{\sc rmfgen} and {\sc arfgen}. 
The PN spectrum is binned to give a minimum of 20 counts per bin.\\
\indent Data from the Reflection Grating Spectrograph 
(RGS; den Herder et al. \cite{den2001}) 
 are also re-processed using the {\sc SAS} {\sc rgsproc} script, 
giving effective exposure time of 19.5\,ks and 18.9\,ks. 
  The signal to noise ratio is not sufficient 
for reliable RGS data analysis.\\
\indent  The OM took a sequence of 4 exposures (1000\,s each) 
through the UVW2 filter. We calculate for all exposures the UV fluxes;
these are, 2.10$\pm$0.04, 2.14$\pm$0.04, 2.11$\pm$0.04, 
and 2.08$\pm$0.04 respectively, expressed in units of 10$^{-14}$\,erg\,cm$^{-2}$\,s$^{-1}$\,A$^{-1}$.
 All values are compatible within the error bars indicating 
no appreciable UV variability.

\section{Spectral energy distribution of Q0056-363:
 bolometric luminosity, and accretion rate}\label{sec:sed}

\begin{figure}[h!]
\psfig{file=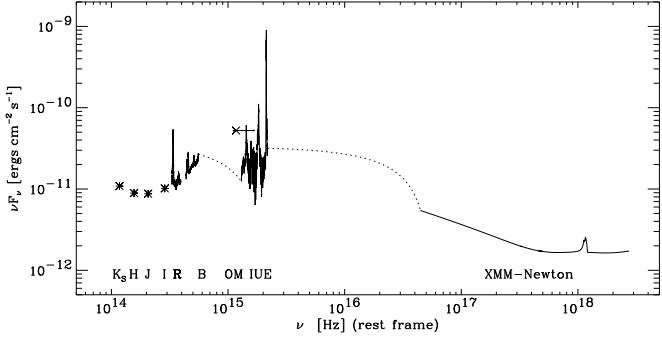,width=8.8cm}
\caption{Spectral energy distribution of Q0056-363. The references 
 of magnitudes, fluxes, and spectra are reported in Table~\ref{table:sed}.}
\label{fig:sed}
\end{figure}

 We use the K$_{\rm s}$, H and J magnitudes from the 2MASS point source 
 catalog released in March 2003 
(Cutri et al. \cite{Cutri2003}, http://www.ipac.caltech.edu/2mass/), 
and the I magnitude from the USNO-B Catalog (Monet et al. \cite{Monet2003}). 
 To convert these magnitude in flux we use the Johnson UBVRI
 photometric system (Campins et al. \cite{Campins85}). 
 We use the optical spectra from Grupe et al. (\cite{Grupe99}, ESO/MPI 2.2m telescope) 
 in the Red and Blue range. The UV spectrum is from an {\sl IUE}  
observation (NASA-ESA, 1999, ftp://archive.stsci.edu/pub/iue/data/).  
 The X-ray part is the unabsorbed {\sl XMM-Newton} PN spectrum (this work). 
 Up to now Q0056-363 is not yet observed above 12\,keV. 
 Table~\ref{table:sed} reports the observation dates and references. 
 We take into account the Galactic extinction 
 using E(B-V)=0.014 mag (Schlegel et al. \cite{Schlegel98}), 
 using the extinction formulae from Seaton (\cite{Seaton79}) for the UV,
 from O'Donnell (\cite{OD94}) for the optical and from Ryter (\cite{Ryter96}) 
 for the IR.  
 The spectral energy distribution for Q0056-363 is reported in 
 Fig.~\ref{fig:sed}.

 Summing the integrated fluxes, from K$_{\rm s}$ and B bands, 
 inside the {\sl IUE} and the {\sl XMM-Newton} 
 domains, we estimate a luminosity of 
 6$\times$10$^{45}$\,erg\,s$^{-1}$. 
 We find that the luminosities in the {\sl IUE} and {\sl XMM-Newton} domains 
 are about 2.0$\times$10$^{45}$\,erg\,s$^{-1}$ and 1.2$\times$10$^{45}$\,erg\,s$^{-1}$, 
respectively.  
 If we take into account the linearly interpolated  fluxes between 
 the {\it B} and the {\sl IUE} domains, 
 and between the {\sl IUE} and {\sl XMM-Newton} spectra (see Fig.~\ref{fig:sed}), 
 we find a bolometric luminosity of about 4.4$\times$10$^{46}$\,erg\,s$^{-1}$. 
 We notice that the largest part of the Eddington luminosity (i.e. $\sim$80$\%$) 
is emitted between the {\sl IUE} and {\sl XMM-Newton} domain, where there is 
no data due to absorption by neutral hydrogen located in our Galaxy, 
 between about 100\,\AA~(120\,eV) and 912\,\AA~ (13.6\,eV).\\ 

\indent According to the relation between the black hole (BH) mass 
and the width and strength of the H$_{\beta}$ line 
 measured in Grupe et al. (\cite{Grupe99}), 
 we infer a BH mass  in Q0056-363 
 of about 4.5$\times$10$^{8}$\,M$_{\odot}$ or of about 
  6.1$\times$10$^{8}$\,M$_{\odot}$ using the 
 formulae from  McLure \& Dunlop (\cite{MD2002}) 
 and McLure \& Jarvis (\cite{MJ2002}) respectively.  
 The Eddington luminosity (L$_{\rm Edd}$=1.26$\times$10$^{38}$~M$_{\rm BH}$/M$_{\odot}$) 
   is then 5.7$\times$10$^{46}$\,erg\,s$^{-1}$ 
   (7.7$\times$10$^{46}$\,erg\,s$^{-1}$)  for M$_{\rm BH}$ =4.5$\times$10$^{8}$\,M$_{\odot}$
  (M$_{\rm BH}$=6.1$\times$10$^{8}$\,M$_{\odot}$). 
 Then the  accretion rate values, as a fraction of the Eddington rate 
 (i.e. $\dot{m}=\dot{M}/\dot{M}_{\rm edd}=L_{\rm bol}/L_{\rm edd}$)  
  are 0.77 and 0.57  for M$_{\rm BH}$ =4.5$\times$10$^{8}$\,M$_{\odot}$ 
 and M$_{\rm BH}$=6.1$\times$10$^{8}$\,M$_{\odot}$, respectively.

\section[]{Spectral analysis}\label{sec:spectra}

We find an average count rate in the 0.3--12\,keV energy range 
 for the PN data  of 2.84$\pm$0.01\,cts\,s$^{-1}$. 
The light curve does not show any evidence of variability, a fit to 
a constant count rate is acceptable
(yielding a $\chi^{2}$ of 148 for 163 degrees of freedom).  
As there is no significant variability detected during this observation, 
we sum the data over the whole observation. 
First, a single absorbed power-law model is fitted to 
the overall 0.3--12\,keV PN spectrum, but we find a poor fit 
 ($\chi^{2}_{\rm red}$= 1.43 for 527 d.o.f. 
with $\Gamma$=2.43$^{+0.02}_{-0.01}$).  
We find no additional absorption, located at the quasar redshift, 
 compared to the Galactic column density 
(i.e., $1.88 \times 10^{20}$\,cm$^{-2}$).
Therefore in all subsequent fits, the column density is fixed to the 
Galactic value.

\begin{figure}[!h]
\resizebox{\hsize}{!}{\rotatebox{-90}{\includegraphics{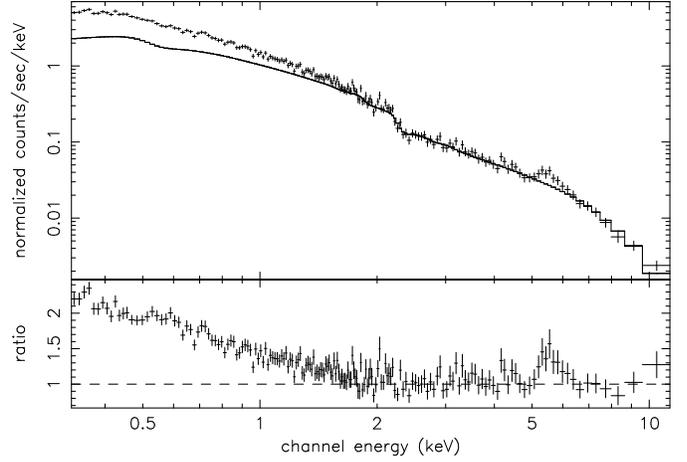}}}
\caption{The PN spectrum of Q0056-363 (in the observer frame). 
A power-law has been fitted to the 2.5--5~keV data
and extrapolated to lower and higher energies. 
A broad, strong soft X-ray excess is
clearly seen extending to $\sim 2$ keV, as well as 
a strong deviation near 5.5\,keV, 
suggesting the presence of a Fe\,K$_{\alpha}$ line. 
For presentation, 
the data have been re-binned into groups of 3 bins, 
after group of a minimum of 20 counts per bin is used for the fit.
}
\label{fig:spectrum}
\end{figure}

To characterize the hard X-ray continuum, we fit an absorbed 
power-law model over the 2.5--5\,keV energy range where the spectrum should
be relatively unaffected by the presence 
of a broad soft excess, a Warm Absorber-Emitter 
medium, an Fe\,K$_{\alpha}$ emission line, 
and of a contribution above 8\,keV from 
a high energy Compton reflection hump.
In this energy range, the data are well fitted by a single power-law model
 with  $\Gamma$= 2.03$\pm$0.18  ($\chi^{2}$/d.o.f.=90.3/98).  
Fig.~\ref{fig:spectrum} displays the spectrum 
extrapolated over the 0.3--12\,keV broad band energy. 
A strong positive residual is clearly seen below 2\,keV 
due to the presence of a soft X-ray excess. 
 In addition, a strong deviation near 5.5\,keV (6.4 keV in the quasar rest-frame) is seen, 
suggesting the presence of a Fe\,K$_{\alpha}$ line complex.

\subsection{The soft excess below 2\,keV}\label{sec:softexcess}
Next, we fit the 0.3--2\,keV energy range with a
single absorbed power-law, and find a reasonable
fit with a photon index of $\Gamma$=2.53$\pm$0.02
 ($\chi^{2}$/d.o.f.=338.8/327). 
 This value is compatible with 
 $\Gamma$=2.7$\pm$0.2 (Grupe et al. \cite{Grupe99}) 
 and  with the slope between 
 2\,500\AA~ and 2 keV of $\Gamma$=2.62$\pm$1.51 
 inferred from December 1990 observation 
  ({\sl ROSAT} All-Sky Survey,  Grupe et al. \cite{Grupe2001}).  
The 0.3--5\,keV energy range is well fitted by the combination 
of a blackbody and a power-law component. 
We find kT$_{\rm bb}$= 133$\pm$7\,eV 
and $\Gamma$=2.28$\pm$0.02 ($\chi^{2}$/d.o.f.=505.7/474).
Since the soft-excess feature is both large and broad, 
we add two other blackbody
components and a better fit is obtained:
kT$_{\rm bb1}$=30$^{+3}_{-2}$\,eV,
kT$_{\rm bb2}$=122$\pm$4\,eV,
kT$_{\rm bb3}$=273$\pm$10\,eV,
 and $\Gamma$=1.93$^{+0.09}_{-0.02}$ ($\chi^{2}$/d.o.f.=472.7/471).  
The three blackbody components represent a large percentage 
(compared to the power-law component) of the total
flux in the 0.3--2\,keV energy range, of 51$^{+47}_{-3}\%$.
The high-energy power-law slope in the multi-component model is consistent
with the $\Gamma$=1.9 index commonly seen in radio-quiet 
quasars (e.g. Reeves \& Turner \cite{R2000}). 
In order to obtain a more physical representation of the soft excess, we 
also test multi-temperature disc models, which may be expected if the 
soft X-ray excess originates via thermal emission from the inner accretion 
disc in Q0056-363. The {\sc diskbb} (non-relativistic) and {\sc diskpn} 
(relativistic) models within XSPEC are used, together with a power-law to 
model the hard X-ray emission above 2 keV. 
Equally good fits are obtained for both models. 
 For the first model we find kT$_{\rm diskbb}$= 161$\pm$9\,eV 
and $\Gamma$=2.21$\pm$0.05 ($\chi^{2}$/d.o.f.=495.6/474), 
 and similar parameters are found for the second model
 with  kT$_{\rm diskpn}$= 153$^{+7}_{-8}$\,eV 
and $\Gamma$=2.21$^{+0.05}_{-0.02}$ ($\chi^{2}$/d.o.f.=495.4/474). 
 The inner disc temperatures obtained through either of these models 
 appear to be unusually high, for what one would expect from a 
  standard steady state $\alpha$ thin accretion disc.  
 Assuming a BH mass of about 4.5--6$\times$10$^{8}$\,M$_{\odot}$, 
  we expect a maximum temperature of only about 15\,eV 
 at 3\,R$_{\rm S}$ (e.g, Peterson \cite{P97}). \\
\indent  We find no evidence for significant intrinsic cold or warm
absorption. Adding two absorption edges of {\ion{O}{vii}} (0.7\,keV) 
or {\ion{O}{viii}}  (0.87\,keV), 
we infer edge optical depth limits of $\tau <0.1$. 
 The lack of significant spectral absorption 
features implies that we are seeing the bare quasar continuum emission.

\subsection{The Fe\,K$_{\alpha}$ line near 6.4\,keV}

As shown in Fig.\ref{fig:spectrum},
a very strong deviation is seen near 5.5\,keV in the observer frame,
i.e. about 6.4\,keV in the quasar frame. 
In the overall 2.5--12\,keV energy band, 
adding a Gaussian line to a single power-law model 
drops the $\chi^{2}$ value by 17 with the addition of 3 degrees of freedom 
(Table~\ref{tab:line}), significant at $>$ 99.95$\%$ according to the F-test. 
Adding an ionised emission line and an absorption edge do not
improve the fit ($\Delta\chi^{2}<$2 for 4 additional parameters).

\begin{table}[!ht]
\caption{Best-fitting spectral parameters in the 2.5-12\,keV energy range
 for an absorbed ({\cal N}$_{\rm H}$=1.88$\times$10$^{20}$\,cm$^{-2}$) 
power-law (PL) component plus a line profile: 
GA: Gaussian profile; and {\sc diskline} and {\sc laor}: profile line
emitted by a relativistic accretion disk for a non-rotating BH and 
 a maximally rotating BH, respectively (Fabian et al. \cite{Fabian89}, Laor \cite{Laor91}). 
The line fluxes are expressed in erg\,cm$^{-2}$\,s$^{1}$. 
 We assume an emissivity law $q$ equal to -2.
 (a): R$_{\rm in}$=6\,R$_{\rm g}$ and R$_{\rm out}$=1\,000\,R$_{\rm g}$, inclination=30$^{\rm o}$.
 (b): R$_{\rm in}$=1.26\,R$_{\rm g}$ and R$_{\rm out}$=400\,R$_{\rm g}$, inclination=30$^{\rm o}$. 
(c): The energy of the line has been fixed to 6.4\,keV (see text). 
}
\begin{tabular}{llll}
\hline
\hline
\noalign {\smallskip}                       
{\small Model}     &    {\small $\Gamma$}            & {\small Line parameters }                          &  {\small $\chi^{2}$/d.o.f.}   \\
\noalign {\smallskip}                       
\hline
\noalign {\smallskip}                       
{\small PL }                       & {\small 1.98$\pm$0.08} &                                              & {\small 152.1/150} \\ 
\noalign {\smallskip}                       
\hline
\noalign {\smallskip}                       
{\small PL}          & {\small 2.03$\pm$0.09}    & {\small E=6.37$^{+0.18}_{-0.13}$\,keV } & {\small 134.9/147} \\ 
\noalign {\smallskip}                       
{\small + GA}                            &                     & {\small $\sigma$=0.23$^{+0.19}_{-0.11}$\,keV}   &           \\
\noalign {\smallskip}                       
                           &                                          &  {\small F=9.6$\pm$4.0$\times$10$^{-6}$}             &           \\
\noalign {\smallskip}                       
                           &                                          &  {\small EW=275$^{+164}_{-113}$\,eV}             &           \\
\noalign {\smallskip}                       
\hline
\noalign {\smallskip}                       
{\small PL } & {\small 2.03$\pm0.09$} & {\small F=8.4$\pm$3.6$\times$10$^{-6}$} & {\small 137.2/149}  \\
\noalign {\smallskip}                       
{\small + {\sc diskline$^{(a,c)}$}}          &                        & {\small EW=292$\pm$124\,eV}                & \\  
\noalign {\smallskip}                       
\hline
\noalign {\smallskip}                       
{\small PL} & {\small 2.02$\pm0.09$} & {\small F=6.7$\pm$3.6$\times$10$^{-6}$}  & {\small 135.9/149 } \\
\noalign {\smallskip}                       
 {\small + {\sc laor}$^{(b,c)}$}      &                        &  {\small EW=229$\pm$94\,eV }              & \\  
\noalign {\smallskip}                       
\hline
\hline
\end{tabular}
\label{tab:line}
\end{table}
We find that the line is well resolved 
 with a full width at half maximum  (FWHM) velocity width 
of about 25,400\,km\,s$^{-1}$,
 and a large equivalent width of about 275\,eV (Table~\ref{tab:line}). 
The line width indicates that the X-ray emission is 
originating from a region close to the BH in Q0056-363. 
For a Keplerian disc, inclined at 30 degrees 
to the line of sight, this velocity implies that the iron line emission is 
occurring at a typical distance of 30 gravitational radii 
($30R_{g}$) from the putative massive BH.
 The FWHM of the line is only about four times 
 smaller than the one found in the Seyfert\,1 \object{MCG-6-30-15},
 which shows the most extreme broad Fe\,K$_{\alpha}$ line observed up to now
(e.g., 100,000\,km\,s$^{-1}$, Tanaka et al. \cite{Tanaka95}; 
Wilms et al. \cite{Wilms2001}, Fabian et al. \cite{Fabian2002}, Lee et al. \cite{Lee2002}).
 Since the line profile appears broad, and is likely to originate within 
$30R_{g}$ of the BH, we proceed to fit the line with a 
profile expected from a relativistic accretion disk around a non-rotating 
 (Schwarzschild) BH, using the {\sc diskline} model in {\sc xspec} 
from Fabian et al. (\cite{Fabian89}). We find that such a profile,
 with a typical inclination of 30$^{\rm o}$ for a type 1 AGN, 
emitted at a rest-frame energy of 6.4 keV,  
provides an excellent representation of the line observed in Q0056-363 
(see Table~\ref{tab:line}).
An equally good fit is obtained for a maximally rotating BH (Kerr)  
 disc emission line model ({\sc laor}; Laor \cite{Laor91}). 
A higher signal to noise ratio spectrum is required to discriminate 
between the  Schwarzschild and the Kerr BH, and to determine the BH spin, if any.\\
 \indent 

The line profile and intensity are unusual for such a high X-ray luminosity AGN 
 (i.e. L$_{\rm X}>10^{45}$\,erg\,s$^{-1}$),
where any broad line component is generally 
expected to be weak and highly ionised.  
As reported by Nandra et al. (\cite{Nandra97}),
in a composite ASCA spectrum of high luminosity AGN, a very weak or negligible
red wing is found in quasars whilst a blue (or ionised) side may be seen.
Therefore the Fe\,K$_{\alpha}$ line profile in Q\,0056-363 
appears different from the average profile
for such high luminosity objects, as the  
profile is consistent with a line emitted
by a cold material in a relativistic accretion disk. 
The equivalent width of the line is also much 
stronger than that usually found in high luminosity quasars, 
where the overall strength of the line is thought to diminish with luminosity 
(e.g. Reeves \& Turner \cite{R2000}). 
The detection of broad iron lines appears to be rather rare for 
luminous quasars.
We note that a weak, but broad iron line has been reported in 
the radio-loud quasar \object{3C\,273}
(Yaqoob \& Serlemitsos \cite{YS2000}), from {\it ASCA} and {\it RXTE} 
observations.  
However any broad Fe\,K$_{\alpha}$ line profile present in a long (100 ksec) high 
signal to noise {\sl XMM-Newton} observation of 3C\,273  is very 
weak, at least down to the level of the systematic calibration 
uncertainties present in the PN detector, i.e. $<$5$\%$ of the 
continuum level at 6 keV (Reeves \cite{R2002}).
A broad iron line has recently been reported in the 
gravitationally micro-lensed quasar QSO~2237+0305 
(Dai et al. \cite{Dai2003}), from a {\sl Chandra} ACIS observation. However 
it is likely that the (uncertain) intrinsic X-ray luminosity 
of this object is below 10$^{45}$~erg~cm$^{-2}$~s$^{-1}$. 
Broad ionised lines, possibly 
originating from a highly photoionised disc, have been 
observed in the high luminosity AGN, Mrk 205 and Mrk 509
(Reeves et al. \cite{R2001}, Pounds et al. \cite{Pounds2001}),
although both of these objects are a factor of 5-10 lower in X-ray 
luminosity than Q 0056-363. 
 Q0056-363 is presently the highest luminosity   
{\it radio-quiet} quasar that exhibits such an intense and {\it broad}  
Fe\,K$_{\alpha}$ line profile from low ionization iron.  
 
\subsubsection{Can X-ray disc reflection explain the Fe\,K$_{\alpha}$ line?}

\begin{figure}[!h]
\resizebox{\hsize}{!}{\rotatebox{-90}{\includegraphics{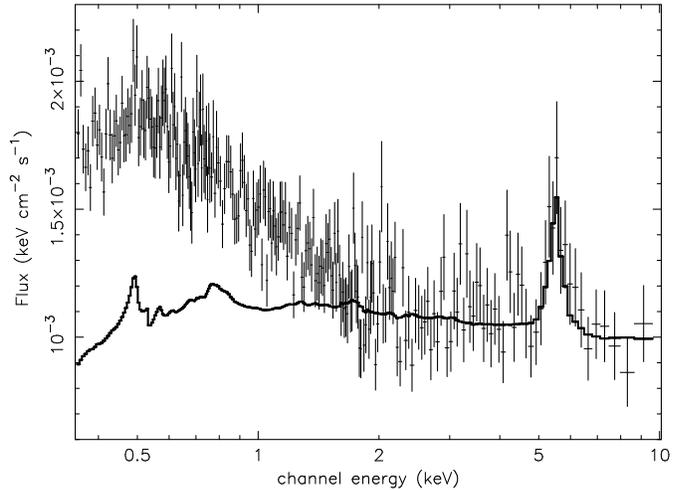}}}
\caption{Ionised disc reflection model fit to the PN spectrum of 
Q0056-363. This model (solid line) provides an excellent representation 
of the 2-12 keV spectrum and of the Fe\,K$_{\alpha}$ line profile. However the 
reflection model does not explain the soft X-ray excess observed 
in Q0056-363 below 2~keV.}
\label{fig:disc}
\end{figure}

Given the evidence for a broad iron line, where the emission is likely to 
originate from the inner accretion disc, we attempt to fit the 
X-ray spectrum of Q0056-363 with a disc reflection model. We use the 
ionised disc reflection model ({\sc xion}) of Nayakshin et al. (\cite{N2000}), 
in the most simple configuration where the X-rays are emitted in a ``lamppost'' 
geometry at a height of $20R_{g}$ above the accretion disc. 
A ratio of X-ray to disc flux of 0.2 is assumed, 
appropriate for the lamppost geometry, 
as well as a high energy cut-off of 100 keV. 
This model provides an excellent fit above 2 keV 
($\chi^{2}/d.o.f=133.9/147$), matching the line profile 
of Q0056-363 very well.  
 The best-fit of this lamppost ionised reflection model is extrapolated 
down to 0.3 keV and shown in Fig.~\ref{fig:disc}. 
 Whilst the model provides an 
excellent fit at high energies, the model fails to account 
for the strong soft X-ray excess observed below 2 keV, as there is not sufficient continuum 
curvature in the reflection model to account for the excess flux. Thus we 
can rule out disc reflection as the cause of the soft excess in Q0056-363. 
The derived parameters for the fit are:  
$\sim$25$R_{g}$ for the inner disc radius (the outer radius is fixed at 
 1000$R_{g}$), and $\sim15^{\rm o}$ for the disc inclination. 
 We notice that the Fe abundance relative to Solar must be fixed to 5 
in order to account for the high flux of the line. 
   Interestingly a formal upper-limit is derived for the accretion rate of about 0.05  
times the Eddington accretion rate. 
This low value of the accretion rate is being driven by the low value of the line energy (i.e. 
6.4~keV) and the high equivalent width of the line. If one experiments 
with this model by increasing the accretion rate through the disc, then the 
line generally becomes more ionised (increasing to 6.7 keV), 
resulting in a worse fit.  
This is due to the formation of a highly ionised 
layer (dominated by He-like Fe) 
on the disc surface (e.g. Nayakshin \& Kallman \cite{N2001}) at 
higher accretion rates. 
 We then relax the assumption that the X-ray emission originates in a simple 
lamppost geometry. A scenario whereby the emission arises through magnetic 
flares above the disc surface is adopted, the principle difference being 
that the X-ray emission occurs close to the surface of the disc (typically 
1 $R_{g}$), resulting in the ratio of local X-ray flux to disc 
flux being substantially higher. An equally good fit is obtained 
for the line profile ($\chi^{2}$/d.o.f = 134.7/147), with very similar 
parameters to those above. 
 Again a low accretion rate ($<5$\% of Eddington accretion rate) 
 is required to match the energy and 
strength of the line; increasing the accretion rate in the flare model 
leads to the formation of a deep, {\it fully ionised} layer of iron at the 
disc surface, weakening the iron line drastically. \\
 \indent The upper limit on the accretion rate (i.e. ${\dot{m}}<$5$\%$) 
 for the lamppost and the magnetic flare models is much lower than the 
 value ($\sim$ 0.6--0.8) inferred according to the spectral energy distribution 
 of Q0056-363 (see sect.~\ref{sec:sed}).  
 This upper limit would be compatible with a much higher BH mass 
 of about 7$\times$10$^{9}$\,M$_{\odot}$. 
 Another possible alternative is a geometry whereby the X-ray source(s) is 
no longer point-like, but exists in a corona which covers most of 
the inner disc surface. In this scenario, the additional weight of the 
corona will increase the gas pressure and density at the surface layers 
of the disc, suppressing formation of a highly ionised layer, 
the result being that the gas remains cool and a line at 6.4 
keV is observed (Nayakshin \& Kallman \cite{N2001}). The energy 
of the line in this case is largely independent of the accretion rate, 
presenting a plausible explanation of the iron line profile 
observed in Q0056-363. 
 For a corona covering most of the disk, one would
 expect a large X-ray flux to UV flux ratio, 
but as is found in sect.~\ref{sec:sed}, 
 the luminosity in the {\sl IUE} domain is about two times higher 
than in the {\sl XMM-Newton} band. However a patchy corona, 
covering a large part (but not all) of the disc surface, 
could give a consistent explanation of the data. \\
\indent  A variability study of the line flux, and whether it is 
correlated or not 
 with the continuum variability, is of great interest and would 
provide a future test (in a longer observation) to discriminate 
 between the above models.  
 As discussed by Reynolds et al.  (\cite{Reynolds99}), 
 the temporal response of the line contains important information
 on the accretion disk structure, the X-ray source geometry and 
 on the BH spin. 
  In the scenario of a dramatic flare in the disc corona, 
  the intensity of the Fe\,K$_{\alpha}$ line is expected 
 to be constant even though the continuum flux varies
 significantly for outflowing magnetic flares with 
different bulk velocities (Lu \& Yu \cite{LY2001}), 
while in the disc-corona geometry it is expected that 
 the line responds rapidly to any change of the continuum.

\begin{table}[!h]
\caption{Best-fitting spectral parameters in the 0.3--12\,keV energy range.
{\sc compTT}: Comptonization of soft photons in a hot plasma (Titarchuk \cite{Titarchuk94}), 
and {\sc diskline}: profile line emitted by a relativistic accretion disk around a non-rotating BH 
 (Fabian et al. \cite{Fabian89}).
(a): The soft photon temperature in the harder {\sc compTT} component (kT=100\,keV) 
has been fixed to the plasma temperature of the softer component. 
(b): R$_{\rm in}$=6\,R$_{\rm g}$ and R$_{\rm out}$=1000\,R$_{\rm g}$, inclination=30$^{\rm o}$. 
(c): The line energy has been fixed at 6.4\,keV in the quasar frame. 
(f): frozen parameter (see text).}
\begin{tabular}{lll}
\hline
\hline
\noalign {\smallskip}                       
   {\sc 2  compTT}$^{(a)}$        + {\sc diskline}$^{(b)}$  \\
\noalign {\smallskip}                       
\hline
\noalign {\smallskip}                       
  kT$_{\rm photon1}$=15\,eV (f)             \\
\noalign {\smallskip}            
      kT$_{\rm plasma1}$= 1.24$^{+0.27}_{-0.22}$\,keV ~~~ $\tau_{_{1}}$=6.6$^{+0.5}_{-0.8}$  \\
\noalign {\smallskip}                     
\noalign {\smallskip}                     
kT$_{\rm photon2}$= kT$_{\rm plasma1}$ (tied) \\
kT$_{\rm plasma2}$= 100\,keV (f) ~~~ $\tau_{_{2}}$=0.4$^{+0.5}_{-0.3}$   \\
\noalign {\smallskip}                       
\noalign {\smallskip}                       
 F$_{\rm line}^{(c)}$=8.3$\pm$3.8$\times$10$^{-6}$\,erg\,cm$^{-2}$\,s$^{-1}$ ~~~ EW=287$\pm$130\,eV        \\
\noalign {\smallskip}                       
\noalign {\smallskip}                       
 F$^{0.3-12\,keV}_{\rm cont}$=8.7$^{+2.3}_{-2.0}\times$10$^{-12}$\,erg\,cm$^{-2}$\,s$^{-1}$   \\
\noalign {\smallskip}                       
\noalign {\smallskip}                        
$\chi^{2}$/d.o.f.=531.3/524          \\
\noalign {\smallskip}                        
\hline
\hline
\end{tabular}
\label{table2}
\end{table}

\subsection{Can Comptonization explain the broad soft X-ray excess in 
Q0056-363?}

\begin{figure}
\psfig{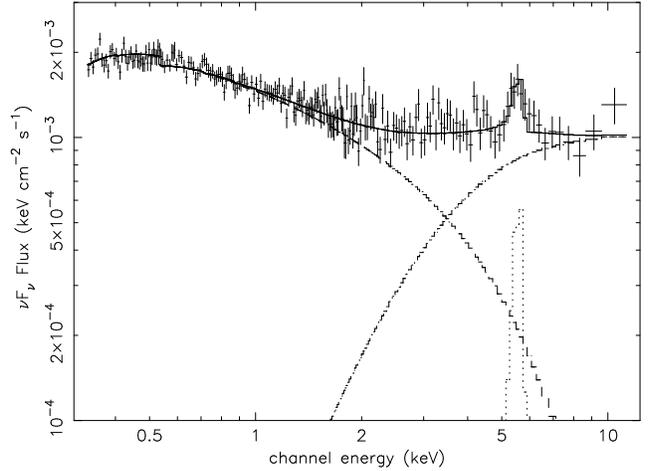}
\caption{The unfolded $\nu F_{\nu}$ PN spectrum (0.3--12\,keV) 
fitted with a double Comptonization model ({\sc compTT}, 
Titarchuk \cite{Titarchuk94})
 and a line profile from a relativistic accretion disk 
({\sc diskline}, Fabian et al. \cite{Fabian89}).}
\label{fig:ufm}
\end{figure}

\begin{figure}
\psfig{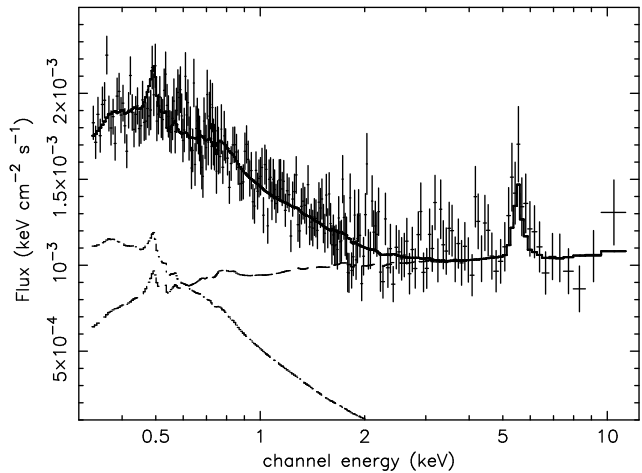}
\caption{The unfolded $\nu F_{\nu}$ PN spectrum (0.3--12\,keV) 
 fitted with a combination of the {\sc compTT} model 
(at soft X-ray energies) and a power-law 
 continuum model which mimics the hard Comptonization component, 
 with an X-ray disc reflection model ({\sc xion}).}
\label{fig:compttxion}
\end{figure}

Comptonization has often been suggested as a source of both the 
soft X-ray and 
hard X-ray spectra of AGN. For example the accretion disc may 
be responsible for the EUV/soft X-ray emission, with some of these 
soft photons being inversed-Compton scattered into the hard X-ray 
energy range, as they pass through the hot corona above the disc.
We further investigate such models over the 0.3-12 keV range,
by using the {\sc compTT} model in XSPEC (Titarchuk \cite{Titarchuk94}).
We first test a model with one absorbed {\sc compTT} component, using  
single electron temperature, plus a {\sc diskline} profile; 
we obtain an unsatisfactory fit with $\chi^{2}$/d.o.f.=664/524, 
 as this model failed to account for all of the soft excess present 
in the X-ray spectrum.  
Instead we test a double Comptonization model, in this model 
we assume that there are two 
layers of Comptonising electrons, whereby the output from the first cooler 
one is in turn Comptonized by the hotter component
 (i.e. kT$_{\rm photon2}$=kT$_{\rm plasma1}$). 
We fix the input soft photons temperature (kT$_{\rm photon1}$)
 of the first {\sc compTT} component to 15\,eV 
 (see sect.~\ref{sec:softexcess} for explanation). 
Since our spectral bandpass (up to 12\,keV) does not enable us to 
constrain the cut-off energy for the hard power-law, the temperature of 
the hotter Comptonization component (kT$_{\rm plasma2}$) has been fixed 
at 100\,keV.
The inferred parameters are reported in Table~\ref{table2}, a very good fit is 
obtained, the model manages to reproduce all of the spectral curvature 
seen in the {\sl XMM-Newton} spectrum of Q0056-363. 
 The largest part of the flux, i.e. about 67$\%$, is emitted in the 0.3--2\,keV energy band. 
As an illustration, Fig.~\ref{fig:ufm} 
shows the corresponding 0.3--12\,keV unfolded 
 PN spectrum (in $\nu F_{\nu}$ units). 
In this model, a hot (or even non-thermal) 
electron plasma is responsible for the power-law like emission 
above 2 keV, whilst a lower 
temperature component is responsible for the soft excess below 2 keV. 
One possibility is that the cooler Comptonizing component  
arises from a hot disc skin or atmosphere whilst the hotter one 
originates from the corona (e.g. produced through magnetic reconnection). 
Alternatively there may be only one Comptonizing layer with a 
non-thermal distribution of electrons (i.e. responsible for the hard X-ray 
power-law), the low energy portion of which 
becomes Maxwellian (i.e. thermalised) and produces the soft excess 
(e.g. see Vaughan et al. \cite{Vaughan2002} for a discussion 
on the soft X-ray excess in the narrow-line Seyfert\,1 \object{Ton\,S180}).\\
\indent While the broad-band X-ray continuum of Q\,0056-363 
can be explained by a double comptonization model, 
this model does not directly account for 
the emission line at 6.4 keV or the reflected continuum from the disc. 
 Therefore we attempted to combine the thermal Comptonization model with 
 the X-ray disc reflection model {\sc xion} 
from Nayakshin et al. (\cite{N2000}),  
which can account for the line at 6.4\,keV. 
The model {\sc xion} can only be directly linked to a power-law 
continuum model, therefore we fit the data with 
a combination of the {\sc compTT} model 
(at soft X-ray energies) and a power-law 
 continuum model which mimics the hard Comptonization component, 
 with an X-ray disc reflection model ({\sc xion}).  
As the previous fits, we fix the disc photon temperature to 15\,eV, 
 and we assume for simplicity a lamppost geometry. 
 We fix the relative abundance of the iron to 5. 
We find a very good representation of both the line at 6.4\,keV 
and the underlying continuum from 0.3 to 12\,keV ($\chi^{2}$/d.o.f.=533.1/522), 
with kT=0.34$^{+0.10}_{-0.07}$\,keV and $\tau$=12.5$^{+2.3}_{-2.0}$
  (for the cool comptonized layer), 
 $\Gamma$=1.98$^{+0.07}_{-0.08}$ (hot comptonized layer), 
 $\dot{m}<$13$\%$,  
 a disc inclination of less than 31$^{\rm o}$,
 and an inner accretion disc radius of 48$^{+73}_{-23}$\,R$_{\rm g}$. 
 We find an unabsorbed 0.3--10\,keV flux 
 of 8.4$\times$10$^{-12}$\,erg\,cm$^{-2}$\,s$^{-1}$ 
and a corresponding 0.3--10\,keV luminosity 
 of 1.2$\times$10$^{45}$\,erg\,s$^{-1}$, 
 with about 67$\%$ is emitted below 2\,keV. 
 The unfolded 0.3--12\,keV unfolded 
 PN spectrum (in $\nu F_{\nu}$ units) is plotted in Fig.~\ref{fig:compttxion}. \\

\section{Conclusion}
We find  that Q0056-363 is a powerful
quasar in the X-ray band, with a 0.3--12~keV unabsorbed luminosity
of about $1.2 \times 10^{45}$ erg s$^{-1}$ with the largest part 
($\sim$ 67$\%$) emitted below 2\,keV.
 The 0.3--12~keV broad band X-ray spectrum of Q0056-363 is dominated 
by a strong soft X-ray excess, and displays a broad and intense Fe\,K$_{\alpha}$ 
line at 6.4\,keV. Q0056-363 is presently the most  luminous AGN known 
to exhibit such a broad and intense Fe\,K$_{\alpha}$ line profile 
 from near neutral iron. 
 The 0.3-12\,keV broad band spectrum of Q0056-363 may be 
 represented by the combination of Comptonization and X-ray disc reflection: 
the cool layer of comptonizing electrons (kT=0.3$-$1.2\,keV)  is then comptonized by the hot
   layer (kT=100\,keV), and the output of that illuminates the disk which emits the line
 at 6.4\,keV. A high relative abundance of the iron of about 5 
 is required to account for the intensity of the line 
 for the disc reflection models. A rather low value for the quasar accretion 
rate (of $<$5$-$13$\%$ of the Eddington rate) is inferred from disc reflection 
models and is not compatible with the rate inferred from the spectral 
 energy distribution of Q0056-363, unless the black hole mass 
 is much higher than the value found according to the relation between 
H$_{\beta}$ line
 widths and flux.   
 One alternative is that the source of X-rays is a patchy corona 
covering a large part of the inner disc surface. \\
\indent  Future high signal to noise observations by 
{\sl XMM-Newton} of Q0056-363 will help to discriminate between a 
 non rotating BH or a rotating BH and to determine the BH spin if any. 
In addition to this, observations  of other bright, high luminosity quasars, 
will help to determine whether the Fe\,K$_{\alpha}$ line in Q0056-363 is truly 
unusual or not. Studying the properties of the Fe\,K$_{\alpha}$ line in 
a variety of AGN, spanning a wide range of physical parameters 
(black hole mass and spin, accretion rate, radio-loudness) 
can provide a potentially powerful diagnostic 
of the accretion process and of the 
geometry of the X-ray emission.

\section*{Acknowledgments}

Based on observations obtained with the XMM-Newton, and ESA science
mission with instruments and contributions directly funded by ESA
member states and the USA (NASA). 
We thank the anonymous referee for very fruitful comments and suggestions.
D.P. acknowledges N. Grosso for his help for the management of 
 the multi-wavelength archived observations of Q0056-363.  
D.P. acknowledges grant support from MPE fellowship.

\end{document}